\documentclass{JINST}
\usepackage{amsmath,amssymb,mathrsfs,graphicx, subfig}

\title{Investigation of background in large-area neutron detectors due to alpha emission from impurities in aluminium}

\author{
J.~Birch$^a$, J.-C.~Buffet$^b$, J.-F.~Clergeau$^b$ P.~van Esch$^b$, M.~Ferraton$^b$, B.~Guerard$^b$, R.~Hall-Wilton$^{c,d}$, L.~Hultman$^a$, C.~H\"oglund$^{a,c}$, J.~Jensen$^{a}$, A. Khaplanov$^{b,c}$\thanks{Corresponding author.}, F.~Piscitelli$^{b,c}$ \\
\llap{$^a$}Link\"{o}ping University, Thin Film Physics division, IFM,
 SE-581 83 Link\"{o}ping, Sweden\\
\llap{$^b$}Institute Laue Langevin,
  71 avenue des Martyrs, FR-38042 Grenoble, France\\
\llap{$^c$}European Spallation Source,
  P.O Box 176, SE-22100 Lund, Sweden\\
\llap{$^d$}Mid-Sweden University,
SE-85170 Sundsvall, Sweden.\\
E-mail: \email{Anton.Khaplanov@esss.se}}

\abstract{
Thermal neutron detector based on films of $^{10}$B$_4$C have been developed as an alternative to $^3$He detectors. In particular, The Multi-Grid detector concept is considered for future large area detectors for ESS and ILL instruments. An excellent signal-to-background ratio is essential to attain expected scientific results. Aluminium is the most natural material for the mechanical structure of of the Multi-Grid detector and other similar concepts due to its mechanical and neutronic properties. Due to natural concentration of $\alpha$ emitters, however, the background from $\alpha$ particles misidentified as neutrons can be unacceptably high. We present our experience operating a detector prototype affected by this issue. Monte Carlo simulations have been used to confirm the background as $\alpha$ particles. The issues have been addressed in the more recent implementations of the Multi-Grid detector by the use of purified aluminium as well as Ni-plating of standard aluminium. The result is the  reduction in background by two orders of magnitude. A new large-area prototype has been built incorporating these modifications.
}


\begin{document}

\section{Introduction}

Large-area detectors for thermal and cold neutrons are required in neutron scattering instruments such as Time-of-Flight (ToF) spectrometers. A number of instruments of this type have been built at reactor and spallation-based neutron sources. A few examples of such present day flagship instruments are IN5 at the ILL~\cite{cite:in5}, LET at ISIS~\cite{cite:let}, CNCS at the SNS~\cite{cite:cncs} and 4SEASONS at J-PARC~\cite{cite:4seasons}. New ToF spectrometers such as CSPEC and VOR~\cite{cite:vor} are being built as a part of the instrument suite of the European Spallation Source (ESS)~\cite{cite:cdr, cite:tdr}. A general characteristic of these instruments is a detector array with an area of 20-40$m^2$ used to measure the positions and arrival times of neutrons that have been scattered by a much smaller sample (typically a few $mm$). Neutrons impinge on a sample in pulses and the majority are scattered elastically, therefore also arriving at the detector in a pulse (although with a wide angular spread). A portion of the scattered neutrons, however, exchange energy with the sample, thus either gaining or loosing velocity. These inelastically-scattered neutrons arrive at the detector over the entire period between the original pulses. The distribution of neutrons both in angle and time-of-flight carry information on the properties of the sample%Features such as inelastic peaks or the broadening of the elastic peak carry information on the properties of the sample. %An example of a measurement where counts are plotted as a function of the detection time-of-flight (which corresponds to neutron energy) is shown in figure~\ref{fig:tofspec}.

Scattering rates for the interesting features in a ToF spectrum vary greatly from sample to sample, but can be very low in some cases -- on the order of 1 neutron per second over the entire detector area is not uncommon. In these measurement conditions, the rate of background counts in the detector must be extremely low. While most ToF spectrometers currently in operation use detectors based on $^3$He, the availability of this rare gas is highly unlikely on the scale necessary for future instruments~\cite{cite:he31, cite:he32, cite:he33}. Due to this, new technologies have been developed~\cite{cite:kirstein}. In this paper we focus on investigation of background in the Boron-10-based Multi-Grid detector which is currently the preferred option for the upcoming ToF spectrometers at the ESS. Prototypes of this gas-based detector have been characterised previously~\cite{cite:khaplanov1, cite:correa, cite:correathes, cite:guerard, cite:in6test}. Background due to $\gamma$-rays in this detector has been studied in detail~\cite{cite:khaplanov2} and it has been shown to be clearly distinct from the features presented in this paper. 

\begin{figure}[tbp] 
\centering
\includegraphics[width=0.8\textwidth]{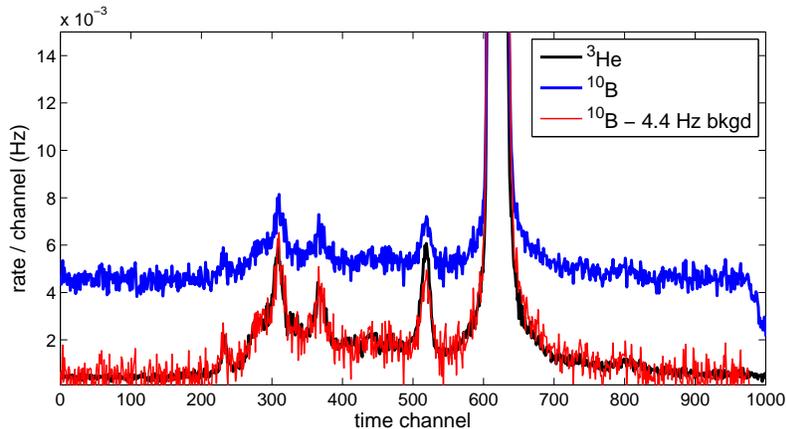}
\caption{Time-of-Flight spectra for the $^{10}$B demonstrator compared to $^3$He detectors during the test on the IN6~\cite{cite:in6test}. Also shown, the ToF spectrum where a constant event rate of 4.4~Hz has been subtracted from the $^{10}$B data. The datasets from the two types of detectors were measured simultaneously by detectors at the same scattering angles, at different heights. Each time channel is a $4.96 \mu s$ interval. The elastic peak tops at about 1000 times higher rate than the background level.}
\label{fig:tofspec}
\end{figure}

While testing a Multi-Grid prototype at the IN6 spectrometer at the ILL~\cite{cite:in6test} a constant background was present that exceeded the background in the original detectors of the instrument by 1-2 orders of magnitude. This difference is shown in figure~\ref{fig:tofspec} (blue and black curves) and it is clear that the signal-to-noise ratio of the new detector needs to be improved. Following the IN6 test, several new prototypes of the Multi-Grid detector have been built, where the background was investigated and ultimately reduced to an acceptable level. In the following, we demonstrate that the background was caused by $\alpha$-particle emissions from the aluminium detector parts, and present steps taken to understand and minimise it.

\section{Origin of $\alpha$ background}

\subsection{The Multi-Grid detector test at IN6}

\begin{figure}[tbp] 
\centering
\reflectbox{\includegraphics[angle=180,origin=c,width=0.37\textwidth]{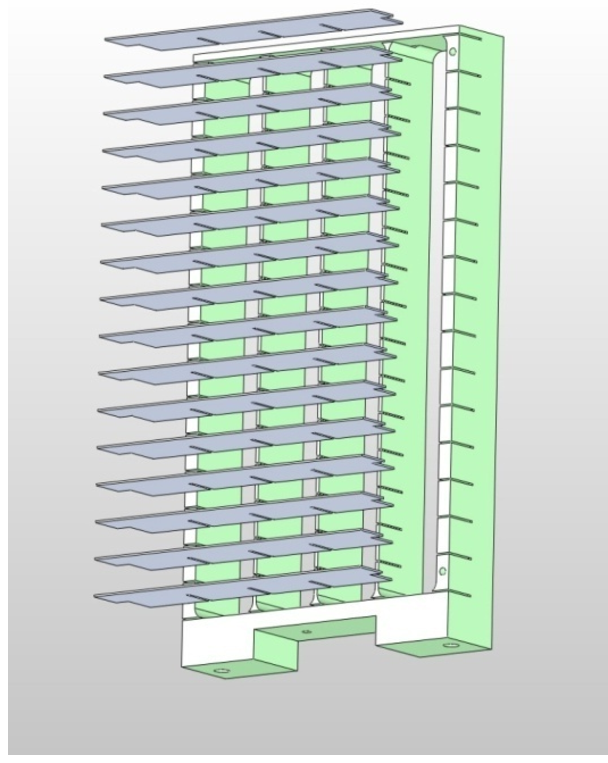}}
\includegraphics[width=0.62\textwidth]{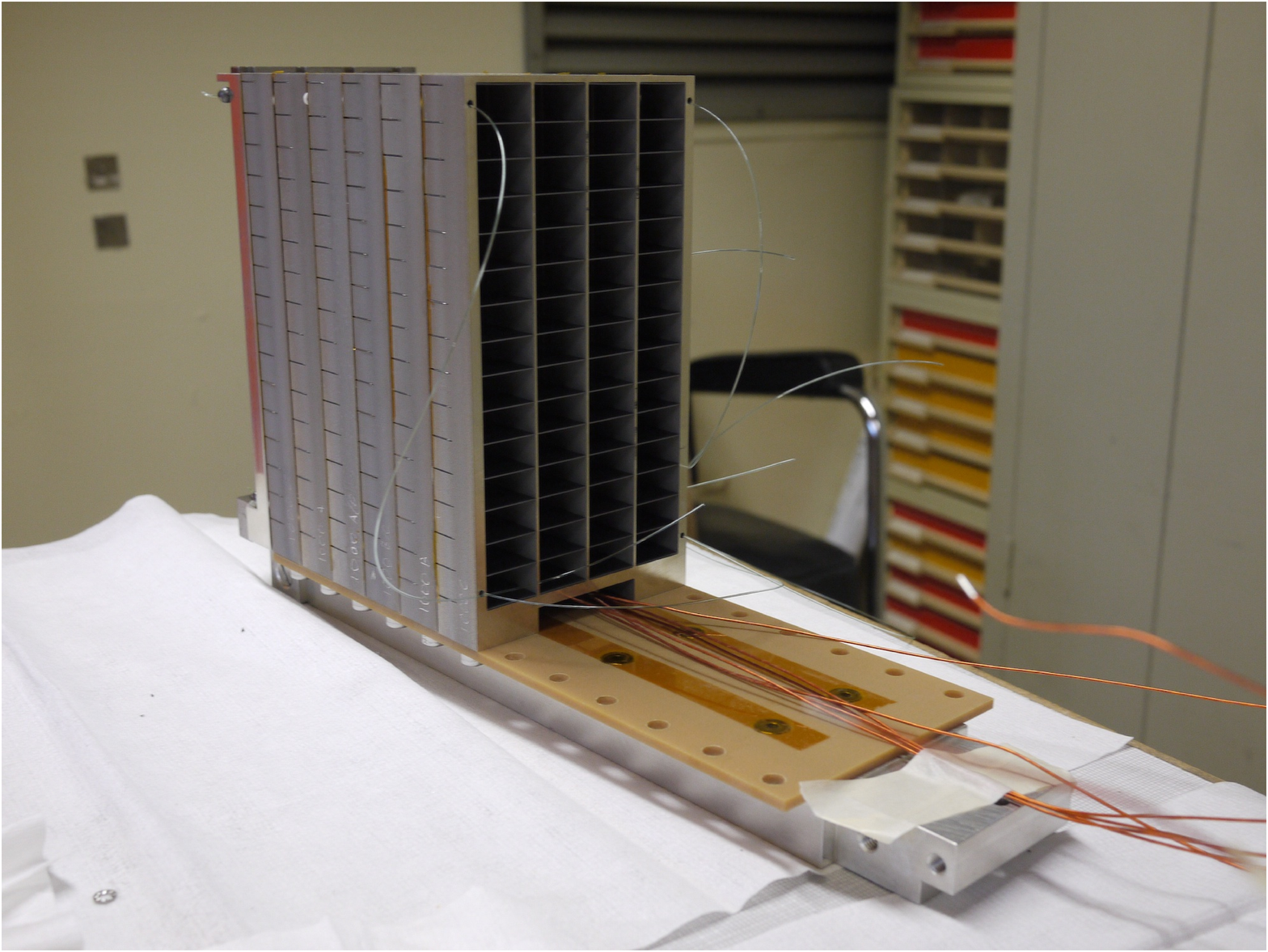}
\caption{\textbf{Left:} A schematic of an aluminium frame with $^{10}B_4C$-coated blades. \textbf{Right:} The 12-frame stack of our small Multi-Grid prototype, partially assembled. This detector was used in many characterisation measurements, and for this work, for measuring $\alpha$-particle emission rate from aluminium components.}
\label{fig:smallproto}
\end{figure}

The Multi-Grid detector is built up of aluminium frames that support 15-20 double layers of $^{10}$B$_4$C coated~\cite{cite:hoglund} onto Al substrates. Coated blades are placed perpendicular to the direction of the incoming neutrons while the uncoated frame elements are parallel to it. Together, they form rectangular gas cells. The dimensions of each cell are typically $20 mm \times 20 mm \times 10mm$, where the $20 mm \times 20 mm$. is coated with $^{10}$B$_4$C. Frames are assembled into a stack as shown in figure~\ref{fig:smallproto}, thus forming rectangular channels that run the length of the stack. Each channel is equipped with an anode wire. Signals are readout from both the anode wires and frames (cathodes) in coincidence providing 3-dimensional position of the interaction to within the size of 1 cell.

\begin{figure}[tbp] 
\centering
\includegraphics[width=0.65\textwidth]{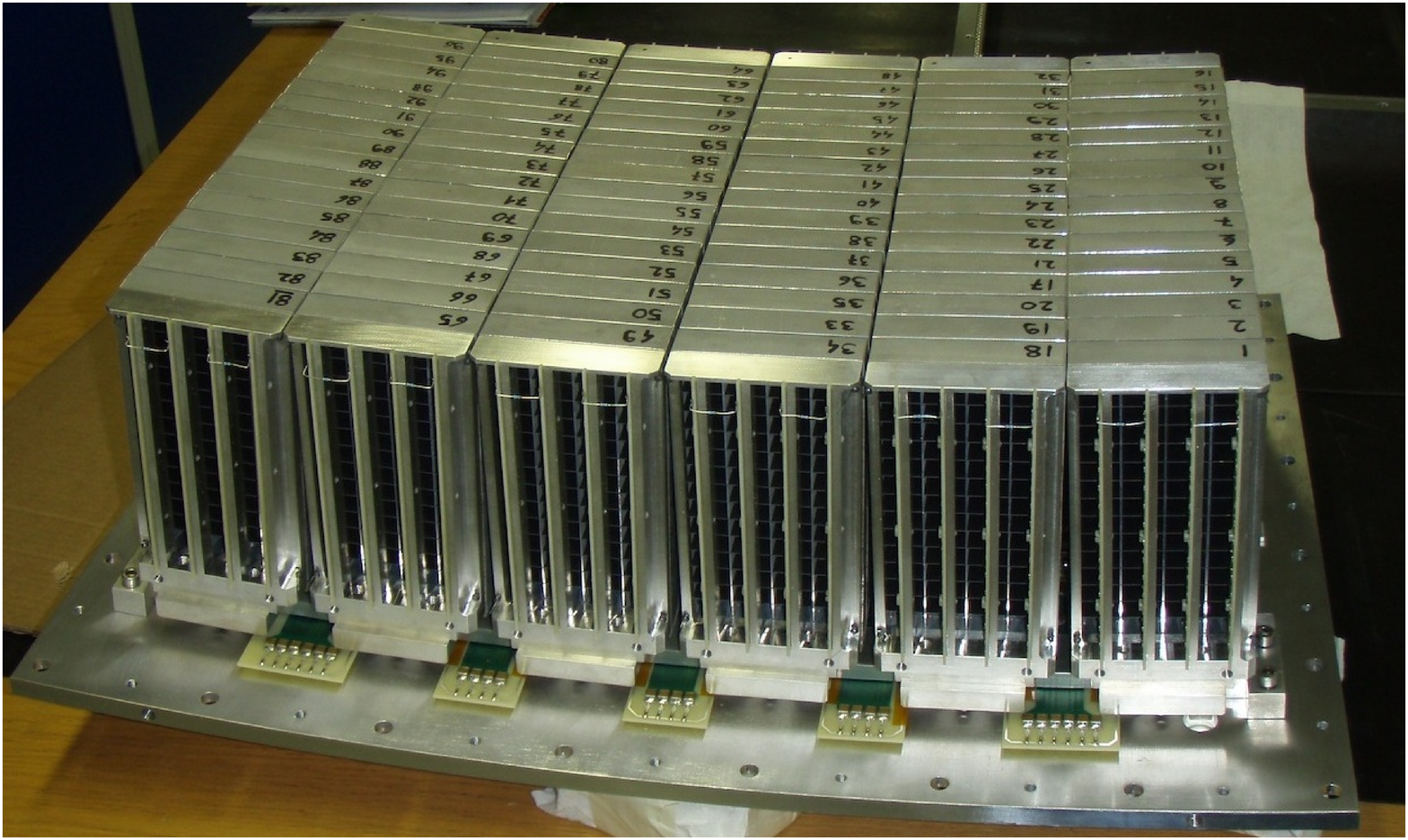}
\includegraphics[width=0.34\textwidth]{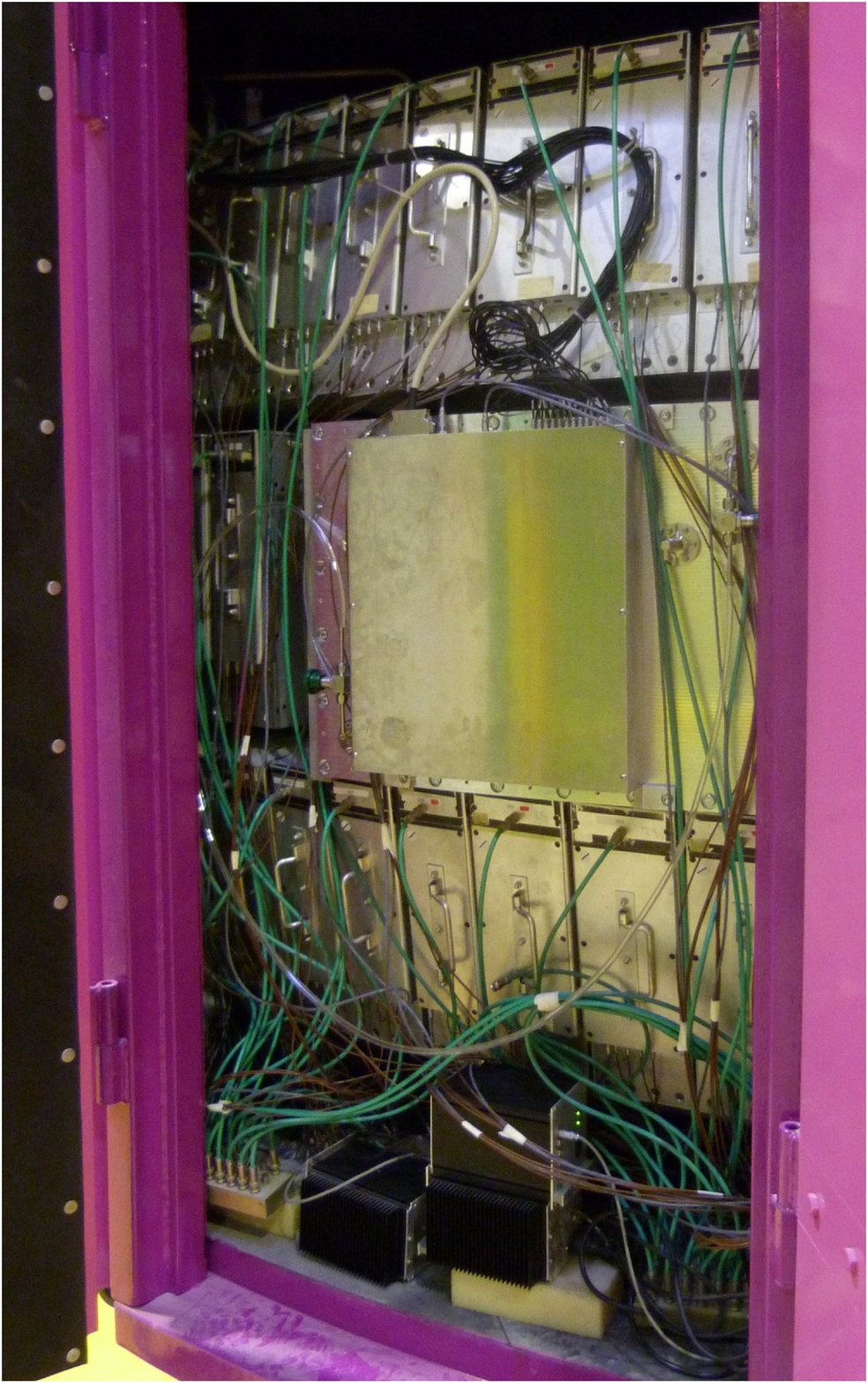}
\caption{\textbf{Left:} The assembly of 6 Multi-Grid stacks used in the IN6 demonstrator. \textbf{Right:} The demonstrator installed at IN6 at the ILL.}
\label{fig:in6test}
\end{figure}

A Multi-Grid demonstrator was built for the IN6 ToF spectrometer at the ILL. Figure~\ref{fig:in6test} shows the demonstrator under construction and installed at the instrument. It is composed of 6 stacks of 16 grids in a curved geometry so that the scattered neutrons are incident normally onto the boron layers. This detector has an active area of $32 \times 50 cm^2$ and it temporarily replaced 25 $^3$He tubes that make up the regular detector bank of this instrument as shown in figure~\ref{fig:in6test} (right). The full description of the results of this test was published previously~\cite{cite:in6test}. Here it is interesting to keep in mind that the total internal surface area of the grids (both B$^4$C-coated and not) that encloses the sensitive gas volume is 6.9~m$^2$ and is composed of aluminium. 

\subsection{$\alpha$ emitters in aluminium}

Aluminium ores tend to contain low concentrations of Thorium and Uranium isotopes. The half-lives of the naturally-occurring $^{232}$Th and $^{235,238}$U are on the order of the age of the Earth, and their concentrations can be considered constant. An $\alpha$ decay leads to much shorter-lived daughter isotopes that $\alpha$- or $\beta$-decay until a stable isotope is reached. 7 or 8 $\alpha$-particles are emitted for each primary decay with energies in the range of 4-9~MeV. The alpha decays found in the decay chains starting with $^{232}$Th and $^{238}$U are shown in table~\ref{tab:chains}.

\begin{table}[tbp]
\caption{$\alpha$-decay energies and half-lives involved in the decay chains starting with $^{232}$Th and $^{238}$U. The most common $\alpha$ energy is given. Branching ratio for $\alpha$-decay is given in parenthesis where it is significantly different from 100\%. $\beta$-decaying isotopes belonging to the chain are also shown for completeness.}
\label{tab:chains}
\smallskip
\begin{tabular}{lcclcc}
\hline
Isotope  		& $\alpha$ energy, MeV		& half-life 				& Isotope 				& $\alpha$ energy, MeV		& half-life \\ 
\hline
$^{232}$Th 	&  4.01 					& $1.4 \times 10^{10} y$	& $^{238}$U  				& 4.20					& $4.67 \times 10^9$ y \\
$^{228}$Ra 	&  $\beta-$				& 5.75 y					& $^{234}$Th  			& $\beta-$ 				& 24.1 d \\
$^{228}$Ac 	&  $\beta-$ 				& 6.13 h					& $^{234}$Pa			  	& $\beta-$  				& 6.70 h \\
$^{228}$Th 	&  5.42	 				& 1.91 h					& $^{234}$U			  	& 4.77  					& $2.45 \times 10^5$ y \\
$^{224}$Ra 	&  5.69 					& 3.66 d					& $^{230}$Th  			& 4.69				  	& $7.54 \times 10^4$ y \\
$^{220}$Rn 	&  6.29 					& 55.6 s					& $^{226}$Ra  			& 4.78  					& 1600 y \\
$^{216}$Po 	&  6.78 					& 0.15 s					& $^{222}$Rn  			& 5.49  					& 3.82 d \\
$^{212}$Pb 	&  $\beta-$  				& 10.6 h					& $^{218}$Po  			& 6.00  					& 3.05 m \\
$^{212}$Bi 	&  6.05 (35.9\%)			& 60.6 m					& $^{214}$Pb  			& $\beta-$ 		 		& 26.8 m \\
$^{212}$Po 	&  8.78 					& 0.30 us					& $^{214}$Bi  				& $\beta-$  				& 19.9 m \\
$^{208}$Tl 	&  $\beta-$ 				& 3.05 m					& $^{214}$Po  			& 7.69  					& 164 us \\
$^{208}$Pb 	&  -- 					&  stable					& $^{210}$Pb  			& $\beta-$  				& 22.3 y \\
  		 	&   	 					&    						& $^{210}$Bi  				& $\beta-$  				& 5.01 d \\
  		 	&   	 					&    						& $^{210}$Po  			& 5.30  					& 138 d \\
  		 	&   	 					&    						& $^{206}$Pb  			& --  					& stable \\
\hline
\end{tabular}
\end{table}

The concentration of uranium and thorium in metallic aluminium and emission rate of $\alpha$-particles from the surface of the metal has been previously studied for the evaluation of the radiological hazard in metallurgy industry~\cite{cite:leroy, cite:hofmann, cite:abbady}, and it is generally found to be well below the level that may cause health concerns. Furthermore, in semiconductor industry, $\alpha$ emissions must be well controlled as they can cause single event upsets in components such as memory~\cite{cite:bouldin, cite:dyer}. While the activity of aluminium is known in the neutron detector community~\cite{cite:zeitelhack, cite:carlile}, few publications exist, as it is only a concern for a small subset of detectors. Concentration of U and Th in aluminium can vary greatly depending on origin; roughly, concentrations on the order of 100 $ppb$ can be expected in standard material~\cite{cite:leroy, cite:hofmann, cite:abbady, cite:bouldin, cite:dyer}. Surface emission rates varying from 0.05 to 0.25~$h^{-1}cm^{-2}$ have been measured for various aluminium samples~\cite{cite:bouldin}. The rate measured in our IN6 demonstrator corresponds to a surface emission rate that falls in this range, closer to the upper end, at 0.21~$h^{-1}cm^{-2}$.

It has also been reported~\cite{cite:leroy} that the concentration of radium isotopes (present in decay chains starting at U or Th), may be different from the equilibrium concentration. This is due to the differences in chemical properties of the elements in the decay chains -- some are transferred directly from the ore to the final product, while others, notably radium, accumulate in the crucibles in the intermediate stages of refinement. The decay of this stored radium contributes additional radioactive isotopes (starting with $^{228}$Ac) years later. The half-lives of the  $^{226}$Ra (part of the $^{238}$U decay chain), $^{228}$Ra (part of the $^{232}$Th decay chain) are 1600 and 5.75 years and so, potentially, the activity of a material can be time-dependent, as well as dependent upon the starting U-to-Th ratio. 

\subsection{Comparison to other background sources}

The background can be identified by considering the energy spectra with and without a neutron flux and the time distribution of the events. The rate of the background was a constant 4.4~Hz in the Multi-Grid IN6 demonstrator while running the experiment, with neighbouring experiments' beams open and closed as well as with the reactor shut down. This leads us to conclude that the background was not caused by neutrons or other particles that may be generated by the reactor or elsewhere in the facility. The rate was also measured in one of the detector laboratories further away from the guide hall. The rate was unaffected by the orientation of the detector, indicating that it is unlikely to be caused by a directionally inhomogeneous radiation such as atmospheric muons, and was unaffected by additional polyethylene and boron shielding, indicating a negligible contribution from atmospheric neutrons. 

\begin{figure}[tbp] 
\centering
\includegraphics[width=0.6\textwidth]{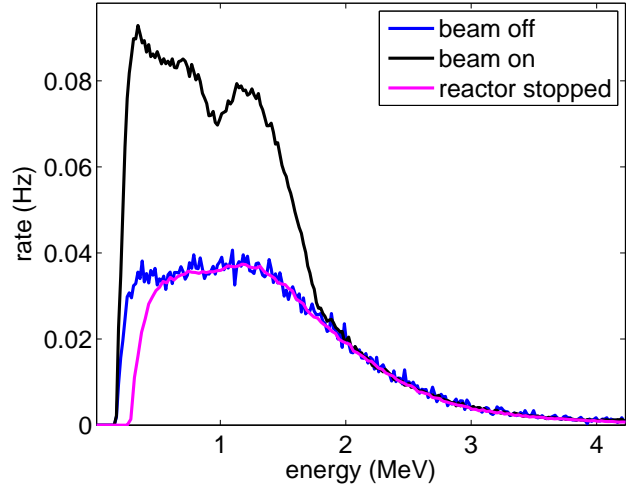}
\caption{Energy spectra measured with and without neutrons present. Rate is presented per energy channel. The energy threshold was changed in the reactor-off measurement resulting in the difference in the low-energy end of the spectrum.}
\label{fig:spect}
\end{figure}

The spectrum of the background events is compared to the spectrum of neutron conversions in figure~\ref{fig:spect}. The neutron spectrum shows the typical two-peak distribution due to the $\alpha$ (up to 1.47~MeV) and $^7$Li (up to 0.84~MeV) detections characteristic of neutron conversions in $^{10}B$. It is superimposed with the background spectrum which extends to higher energies, up to 4~MeV. This latter component is present with no neutrons incident. This spectrum is in stark contrast to spectra produced by $\gamma$-rays, which is found below 100~keV in this type of detector~\cite{cite:khaplanov2}. Similarly to $\gamma$-rays, muons could also produce only low-energy signals, since the energy loss is on the order of few $keV / cm$ for typical energies of atmospheric muons in gas at 1~bar. 

A further source of background could be due to alpha emissions from decay products of radon. Radon isotopes $^{220}$Rn (with 55.6~s half-life) and $^{222}$Rn (with 3.82~d half-life) are produced in the thorium and uranium decay chains respectively. While all other trans-lead elements are solids, radon is a gas, and, as such, can migrate through air before decaying. The solid decay products can be deposited on surfaces, and potentially lead to surface $\alpha$ emissions. This effect would be similar to that of the $\alpha$ emission due to bulk contamination. The decay products of the short-lived $^{220}$Rn are also relatively short-lived, see table~\ref{tab:chains}, and will have decayed within several days after deposition on a surface. The longer-lived $^{222}$Rn, however, leads to $^{210}$Pb, with 22.3~y half-life. This isotope beta-decays and therefore contributes to a background distinct from $\alpha$-particles. It's decay leads to $^{210}$Po, which does $\alpha$-decay with a half-life of 138 days. The $\alpha$ activity of a contaminated surface would therefore increase until $^{210}$Pb and $^{210}$Po are in equilibrium. If the source of radon had been removed, the activity will fall off on a time-scale of decades as $^{210}$Pb is depleted. The effect may further be complicated, as cleaning and surface treatments are more likely to remove $^{210}$Pb than $^{210}$Po~\cite{cite:zuzel}.

The magnitude of radon contamination depends greatly on geographical location as well as on proximity to minerals containing uranium and thorium. Due to the half-lives involved, the history of samples influences the surface $\alpha$ contamination. In order to separate $\alpha$ counts due to radon and bulk contamination, measurements over time are necessary, for samples that have been exposed to radon for a different duration. This type of measurement was not possible in the present study. It is, however, useful to note, that the grids were typically assembled within about 1 month after the production of the parts. Grids were machined from aluminium plate, while blades were chemically etched. Some of the parts were plated with nickel. All of these processes effectively remove the surface of the material. We therefore expect that only the time between production and assembly of the detector, about 1 month, could contribute to contamination. Once assembled, the detectors remained sealed from ambient air. In such conditions, an increase in the $\alpha$ activity may be seen as the decay of $^{210}$Pb populates $^{210}$Po. Systematic measurements of sufficient duration could not be performed, however. 

\subsection{Simulation of the $\alpha$ spectra}

The spectra due to $\alpha$ background in a cell of the Multi-Grid detector have been simulated using GEANT4~\cite{cite:g4}. In order to simulate the contamination of the bulk of aluminium, it was assumed that sources of $\alpha$-particles (radioactive nuclei) are distributed uniformly throughout the aluminium walls of a cell. To simulate radon contamination, a simulation was also made where sources were placed on the surface of aluminium.

\begin{figure}[tbp] 
\centering
\includegraphics[width=0.49\textwidth]{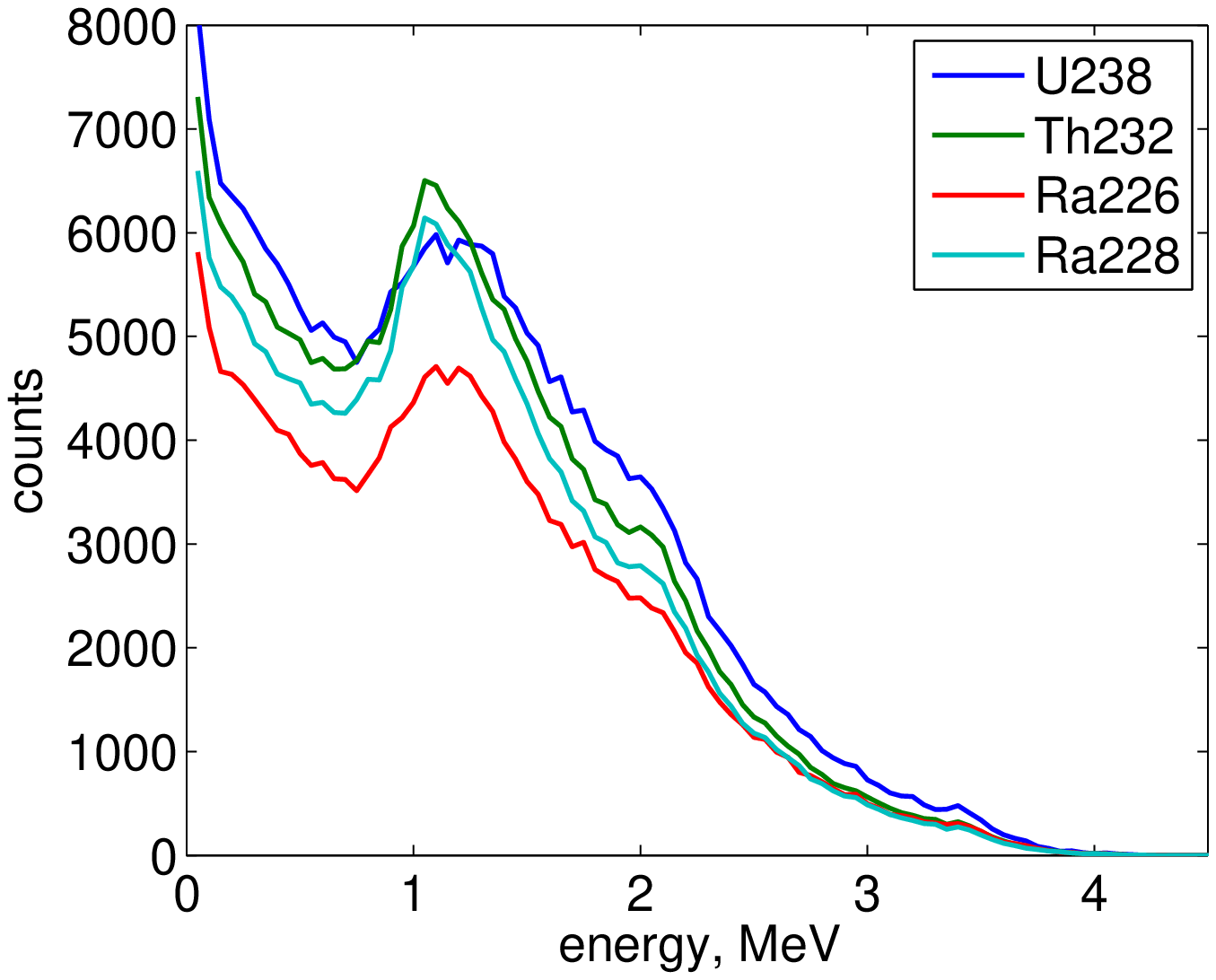}
\includegraphics[width=0.49\textwidth]{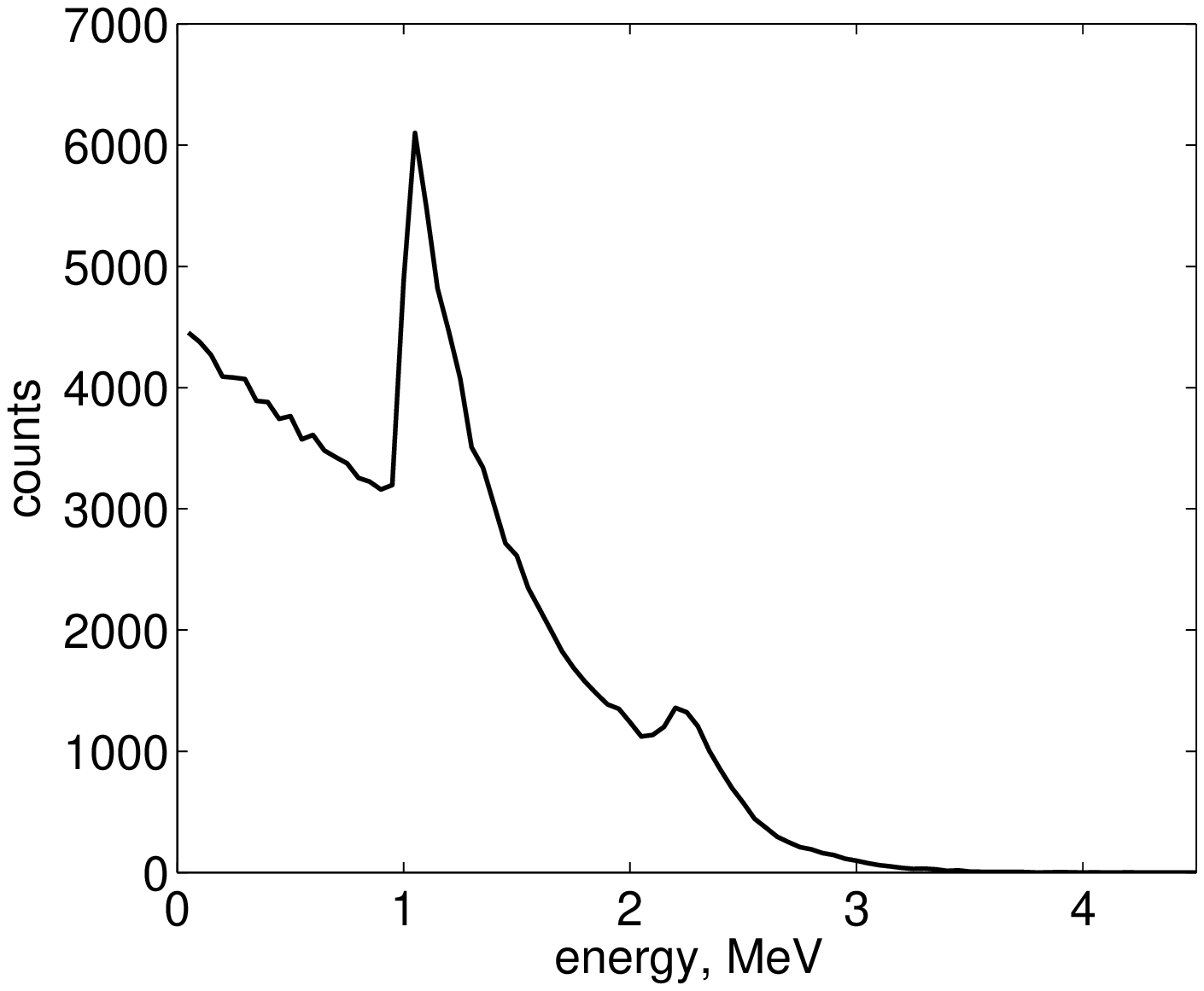}
\caption{Simulated energy spectra of $\alpha$-particles deposited in a cell of the Multi-Grid detector. \textbf{Left:} Spectra due to the decay chains starting with $^{238}$U, $^{232}$Th,$^{226}$Ra and $^{228}$Ra in the bulk of aluminium. Each decay chain is assumed to be in equilibrium. The quantity of the starting isotope is equal for each curve. \textbf{Right:} Spectrum due to $^{210}$Po deposited on the inner surface of a cell.}
\label{fig:simspect}
\end{figure}

The detecting gas was set as a mixture of 900~mbar Ar and 100~mbar CO$_2$. Particles were tracked and energy deposited in gas was considered detected. $\alpha$-particles of several different energies are produced in a decay chain. The isotopes found in the decay chains starting with $^{232}$Th and $^{238}$U are shown in table~\ref{tab:chains} along with corresponding  $\alpha$-decay energies. $\beta$-decays were not included in the simulation, since energy deposited in the detector by $\beta$-particles, \emph{i.e.} electrons, is in another energy range, more similar to that studied in connection to $\gamma$-sensitivity of our detectors~\cite{cite:khaplanov2}. Combining the $\alpha$-particle energies for a particular decay chain, total alpha spectra were obtained. These are shown in figure~\ref{fig:simspect} (left) for decay chains starting at the primordial isotopes, $^{238}$U and $^{232}$Th, and $^{226}$Ra and $^{228}$Ra -- isotopes whose concentration may be enhanced due to the history of a particular batch of aluminium. For the simulation of each decay chain, an equilibrium concentration of the isotopes was assumed. Since we do not have information on relative concentration of U and Th and possible deviation from equilibrium of radium isotopes, we only present the spectra for each isotope separately. For the simulation of radon, only the 5.3~MeV $\alpha$ from $^{210}$Po was considered. The resulting spectrum is shown in figure~\ref{fig:simspect} (right).

The $\alpha$ spectra lack a full-energy peak for two reasons. First, the emission depth is randomly distributed (for bulk emission), and $\alpha$-particles reach the gas after a random loss of energy. Second, many of the $\alpha$-particles that do reach the gas, have sufficient energy to traverse the entire cell and to be absorbed in an opposite wall. Therefore, the shape of the spectra is strongly dependent on the dimensions of the voxels of the Multi-Grid detector. 

The spectra have a peak between 1 and 1.5~MeV and another one, less pronounced, between 2 and 2.5~MeV. Events in these peaks correspond to the most common distances an alpha particle can travel through the cell -- the lower-energy one from a 2-cm wall to the opposite 2-cm wall and the higher-energy one from a 1-cm wall to the opposite 1-cm wall. This effect is very pronounced in the surface spectrum, while in the bulk spectra, the features are smeared due to random depth of emission and a combination of several $\alpha$ energies. The highest energies in the spectrum are those due to the rare trajectories from a corner to the opposite corner. The low-energy parts of the spectra are due to trajectories ending on an adjacent wall, and those escaping the bulk with insufficient energy to traverse the cell. 

The resulting shape and end-point of the spectra is consistent with the spectrum of the background measured in the detector, however the measurements do not show a clear peak as in the simulations. In part, smearing due to energy resolution could be responsible for this. This discrepancy is much larger for the case of the emissions from surface $^{210}$Po contamination than for the case of bulk contamination, suggesting mainly a bulk contribution. 

\section{Background reduction}

The detection of $\alpha$-particles from bulk contamination can be prevented in two ways -- either the $\alpha$-emitter concentration in the aluminium needs to be greatly reduced, or the $\alpha$ particles need to be stopped before entering the sensitive volume. Both of these ideas were applied to the Multi-Grid detector. Specially purified aluminium, where the concentration of $\alpha$ emitters is reduced to below a ppb level, is commercially available (its main application is in semiconductor industry). We purchased 0.5-mm sheet of such radio-pure Al and used it to manufacture the blades. This is a 6000 series aluminium alloy, that, in addition to Al, contains Si at 1\% (also available with 4\% Si) and other elements below 1~ppm. According to the supplier's chemical analysis, the concentration of Th and U are 0.17 and 0.13 ppb (weight), respectively. We also had the grids and blades manufactured from standard alloys (Al5754 and Al5083) and plated with Ni in order to stop the $\alpha$-particles before reaching the detector gas.

We chose to keep aluminium as the main structural material of the detector. The advantages of aluminium are its relatively low neutron absorption and scattering cross-sections, ease of manufacturing and light weight. These factors are considerable for the Multi-Grid detector since large numbers of parts are required so the material budget exposed to neutrons, cost, and total weight are all crucial. In other types of detectors for low background measurements it may be preferable to simply replace aluminium by another metal, possibly accepting a higher neutron cross-section. This is the case for the majority of $^3$He tubes where stainless steel is used. Another alternative, when a thin substrate is mechanically practical, could be a coated foil (of possibly Ni), avoiding Al entirely. 

\subsection{Background tests with a 12-frame prototype}

In order to evaluate the level of alpha emission from aluminium, a large surface area needs to be measured due to the relatively low rate. %Alternatively, a very high-sensitivity measurement is required over a long time, however, this also sets a very strong limit on acceptable level of other background, such as $\gamma$-rays. We considered using a detector with a single large volume gas chamber equipped with a proportional wire readout, where a sheet of aluminium to be investigated could be placed. This method would, however, be limited by $\gamma$-ray counts, since a large gas detection volume allows for $\gamma$-ray signals of larger amplitude than in the small cells of the Multi-Grid due to longer secondary electron track length.
We chose to test different grades of Al as well as Ni-plated Al by building a small Multi-Grid detector using these materials. We used our 12-frame prototype, shown in figure~\ref{fig:smallproto} for this work. This detector has been previously used for characterisation, including efficiency and $\gamma$ sensitivity measurements. While rebuilding the prototype using a new set of components presents a considerable amount of work, it was nevertheless the preferred option as  a set-up for building such detectors in a time-efficient fashion had already been developed. Furthermore, by measuring a new material as a part of a Multi-Grid detector, we are able to compare the findings directly to our previous measurements, as well as evaluate the contribution of $\alpha$ background in the configuration of interest. The total internal area of the prototype (0.86~m$^2$) allowed good statistics to be collected over few-day long measurements. 

Table~\ref{tab:protolist} shows a list of configurations that were built. Both frames and blades contribute to the background. All frames were built using a standard alloy which was Ni-plated in all cases except P11. Blades made from standard Al were used in P11 and P12, pure Al in P14, others were Ni-plated. In most cases, B$_4$C coating was not present in order to be able to measure the $\alpha$ background without any neutron signal. The last detector, P15, included three regions with blades from different suppliers of Ni-plating.  

\begin{table}[tbp]
\caption{Configurations of the 12-frame prototype used for $\alpha$ background evaluation. The achieved background reduction is shown in the last column where the rate in the detector built of standard allow is normalised to 100\%. Ni-plating thickness values are those specified to the supplier -- the results differed, see text.}
\label{tab:protolist}
\smallskip
%\centering
\begin{tabular}{llllr}
\hline
%Energy & \multicolumn{5}{c}{Threshold}\\
 %\cline{2 - 6}
Designation 	& Al (frames) 					& Al (blades) 				& B$_4$C coating 		& $\alpha$ rate, \% \\ 
\hline
P11 			&  Al5083 					& Al5754					&  1 $\mu$m 			& 100 \\
P12 			& 25 $\mu$m elec. Ni 			& Al5754					&  none				& 68.7\\
P13 			& 25 $\mu$m elec. Ni 			& 25 $\mu$m elec. Ni		&  1 $\mu$m			& 26.3 \\
P14 			& 30 $\mu$m chem. Ni 			& pure Al 				&  none				& 1.86 \\
P15  		& 30 $\mu$m chem. Ni 			& 25 $\mu$m chem. Ni	(supplier 1)	&  none				& 2.37 \\
  			& 							& (supplier 2)				& 					& 2.45 \\
  			& 							& (supplier 3)				& 					& 2.23 \\		
\hline
\end{tabular}
\end{table}

\subsection{Investigation of Nickel plating}

Ni-plating of two types and from several suppliers was tested. Assuming a bulk density layer and a zero $\alpha$-emitter content in nickel, the simulation showed that a minimum of 23 $\mu$m thick plating is needed to stop 100\% of the $\alpha$-particles; 10 $\mu$m should stop 95\%. The needed Ni-plating thickness can be achieved by either electrolytic or chemical plating. Both techniques were used.

\begin{figure}[tbp] %\centring
\includegraphics[width=0.49\textwidth]{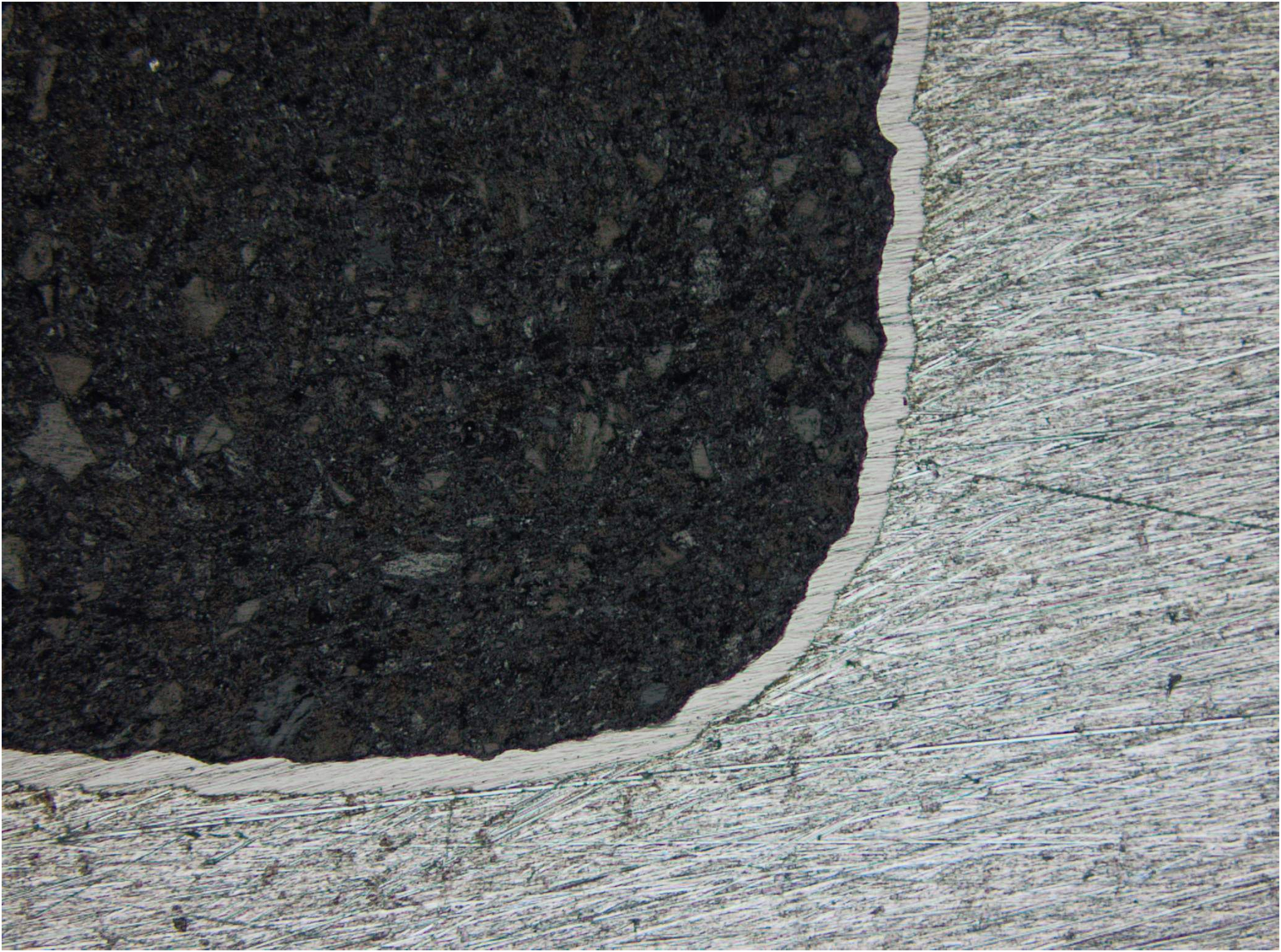} 
\includegraphics[width=0.49\textwidth]{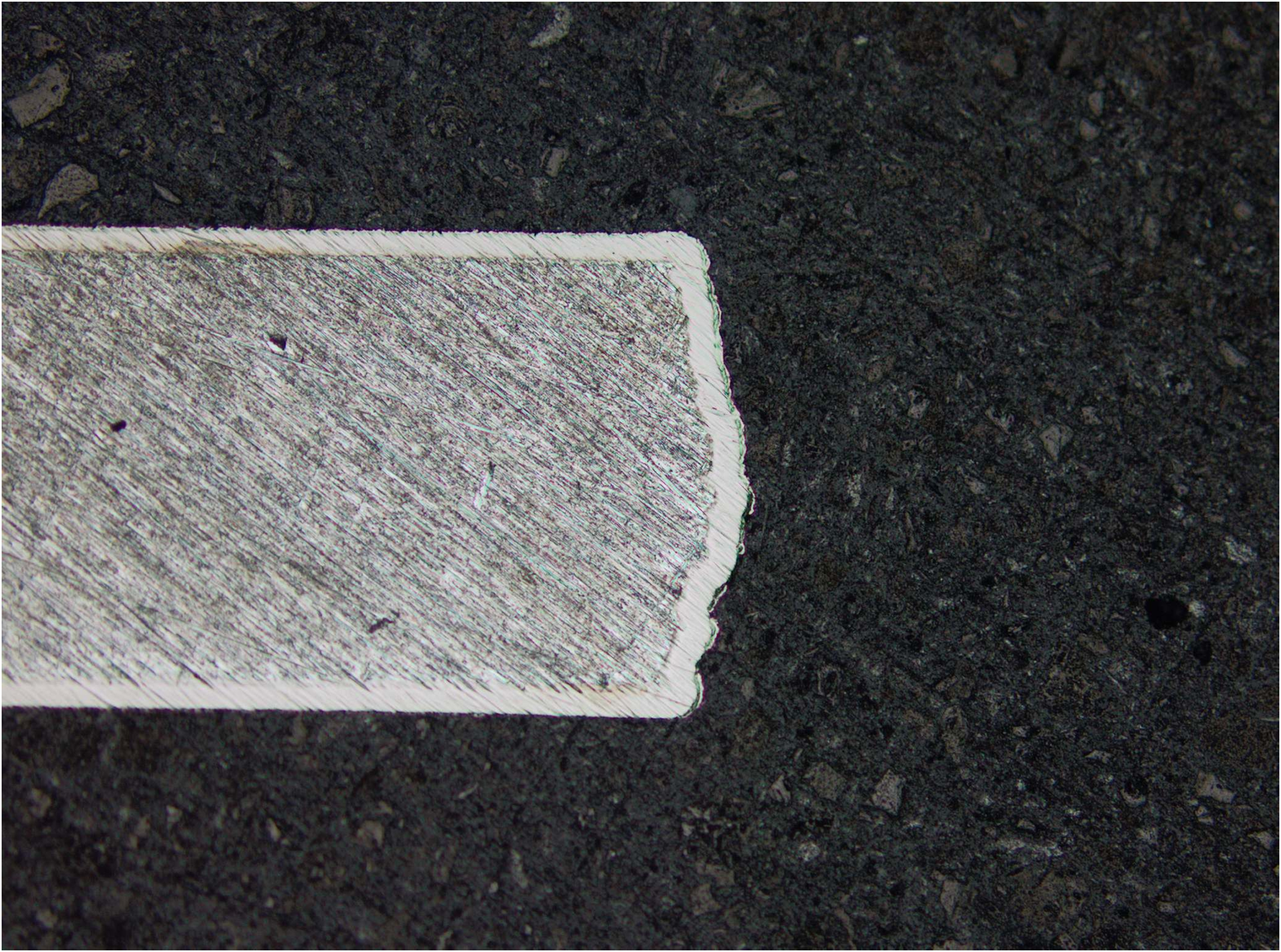} 
\caption{Optical microscope images of a cut cross section of Ni-plated parts encased in bakelite. Chemically Ni-plated grid, used in P14 and P15, (left) and blade, used in P15, (right) are shown. The black material is bakelite, the darker grey -- Al and lighter grey -- Ni. The scale is the same in both images. The thickness of the Al of the blade (right) is 500~$\mu m$}
\label{fig:nithickness}
\end{figure}

The quality of the plating was studied under microscope. Sections of a plated grid or blades were encased in bakelite, cut and polished in order to expose the cross section. Images of the Ni layer taken with an optical microscope Olympus BX51 equipped with a digital CCD camera are shown in figure~\ref{fig:nithickness}. In the left image, a cross section of a grid is shown in the region of one of the corners, and a blade is shown on the right. In both cases, Ni layers produced by chemical plating are shown. It has a very even thickness that varies by at most a few $\mu m$. The layer's surface closely follows the surface of the underlying aluminium, and so appears more even on the blade (which is produced from a rolled and etched sheet), while it is more bumpy on the grid surface which was cut by spark erosion. Chemical plating from 3 suppliers was ordered and used in P14 and P15. The thickness and evenness from each appeared equivalent in microscope investigations, although visually, the layer had a different shade. By contrast, Ni layer produced by electroplating and used in P12 and P13, was much thinner -- while the thickness was specified as 10 (blades) and 25 (blades and grids) $\mu m$, the actual thickness was found to be below 10 $\mu m$ for both batches. This deviation from required thickness is clearly reflected in the much poorer background results from this material. It should be noted that these results are based on only one supplier of electroplating, and it is unclear whether this was a production error or a general characteristic of the process. 

Each end of a stack of grids has a mechanical support to which the anode wires are attached on each end of the stack (these are seen at the front of the detector in figure~\ref{fig:in6test}). These include an aluminium part with a shape similar to that of a grid. These were also Ni-plated, since $\alpha$ particles emitted from these parts may enter the sensitive volume if emitted in that direction. The plating used for the end-grids is of the same type as the regular grids in that iteration of the detector. It should be noted, however, that there is unshielded material other than Al at these positions, in particular PCBs to which the anode wires are attached as well as solder attaching the wires.

Nickel plating produced by a chemical process could be ordered with 5-30\% weight content of phosphorous. We requested the P concentration to be as low as possible from each of the three manufacturers that we used. In order to verify the content, compositional analysis was performed with time-of-flight elastic recoil detection analysis (ToF-ERDA), using a 36 MeV $^{127}$I$^{8+}$ beam at 66 $^\circ$ incidence and 45$^\circ$ recoil scattering angle. The fraction of phosphorous was found to be 5.5, 8.0 and 11\% (weight). The concentrations of elements other than Ni and P was below 0.1\%.

\subsection{Achieved background reduction}

\begin{figure}[tbp] 
\centering
\includegraphics[width=0.7\textwidth]{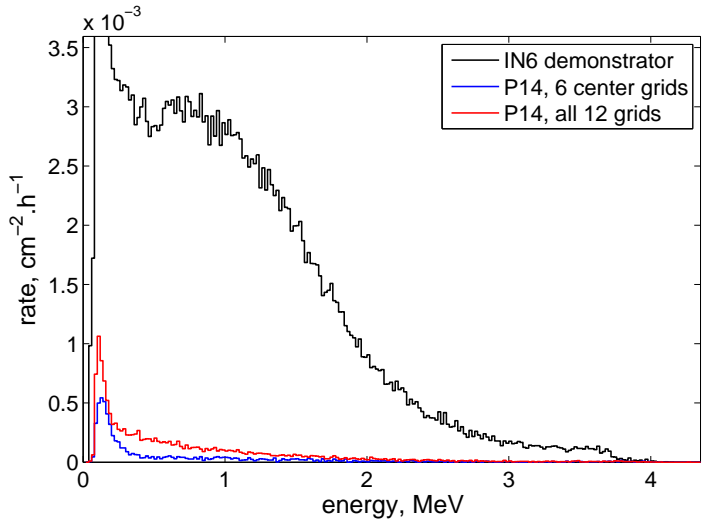} 
\caption{Comparison of the background spectrum in the original detector (IN6 demonstrator) to the background of P14, where the spectra of all and inner-only grids are shown separately.}
\label{fig:result}
\end{figure}

At the conclusion of the measurements with the detectors listed in table~\ref{tab:protolist}, a satisfactory level of background reduction was achieved. This is shown in figure~\ref{fig:result} for P14 (Ni-plated frames and radio-pure Al blades). Very similar result was obtained for P15 (all Ni-plated) for all 3 Ni-plating suppliers. The exact background rates are given in table~\ref{tab:protolist}. The background rate was found to be higher in the grids at the edges of the stack. This indicates that materials outside of the sensitive part of the detector are contributing to $\alpha$ background. While the relevant Al parts were Ni-plated, we suspect that the solder, that was used to attach the anode wires, may be contributing the additional counts. Surface $\alpha$ emission rates on the order of 1 cm$^{-2}$h$^{-1}$ have been reported for tin samples~\cite{cite:gordon}. As tin is the main component in the solder used, this may account for the increased rate in the edge grids. We consider that this effect can be avoided in future designs and therefore will consider the rate in the central grids indicative of the achieved background reduction.

The threshold setting can influence the rates measured. A significant portion (over half) of the rate is seen in the lowest end of the energy spectrum in figure~\ref{fig:result}. It was not possible to determine whether these energies were deposited by alpha or whether this is a contribution of the high-energy tail of backgrounds such as $\gamma$-rays or cosmic muons. While it would be possible to reject these events using a higher threshold, this would also adversely affect the neutron efficiency. Therefore the thresholds were set in the same way that they would be set for a typical neutron measurement with this detector. The thresholds are set above the electronic noise level. The bias voltage is then adjusted so that the gas gain is such that the gamma-ray rejection is also satisfied, as described in our previous publication~\cite{cite:khaplanov2}. The threshold settings and bias voltage were the same in every measurement with the 12-frame prototype. 
%This is based on the requirement of reduction of the $\gamma$-ray background. The exact methodology is described in our previous publication~\cite{cite:khaplanov2}. 

\subsection{Impact on construction of the next Multi-Grid demonstrator}

Following the test of the Multi-Grid on IN6 and the background tests described above, a large-area demonstrator was built. This detector consists of 8 stacks of grids. There are 128 grids (22.5~mm in height) in each stack and 18 blades in each grid. In total, 1024 grids are used to create a detector with a sensitive area that is 3~m long and and 80~cm wide, or 2.4~m$^2$. This size was chosen so that the $^{10}$B detector matches the size of the $^3$He modules currently used in the IN5 instrument at the ILL. While the full description of this detector will be presented elsewhere, we here point out the steps taken to reduce the background due to $\alpha$ events, based on tests presented here.

Both Ni-plating and radio-pure Al were used in the construction of the grids of the IN5 demonstrator. It should be noted that the radio-pure Al alloy that we were able to obtain, 6000 series, is considerably softer than the series 5000 alloys that we previously used for both grids, blades and other parts of the detector. Additionally, the B$_4$C coating process causes the Al to soften further. On the other hand, a 30-$\mu$m layer of Ni plating on each side of a 0.5 or 1-mm thick Al part adds significantly to its rigidity. A combination of Ni-plated grids and pure Al blades resulted in good overall rigidity of the frame and relatively simple assembly. For our IN5 demonstrator, 512 such grids have been successfully assembled. For the other 512 grids of this detector, we used the radio-pure Al for both blades and grids. While these were much more flexible, their rigidity was increased by an additional layer of neutron shielding on the outside of the grid.

\begin{figure}[tbp] 
\centering
\includegraphics[width=0.49\textwidth]{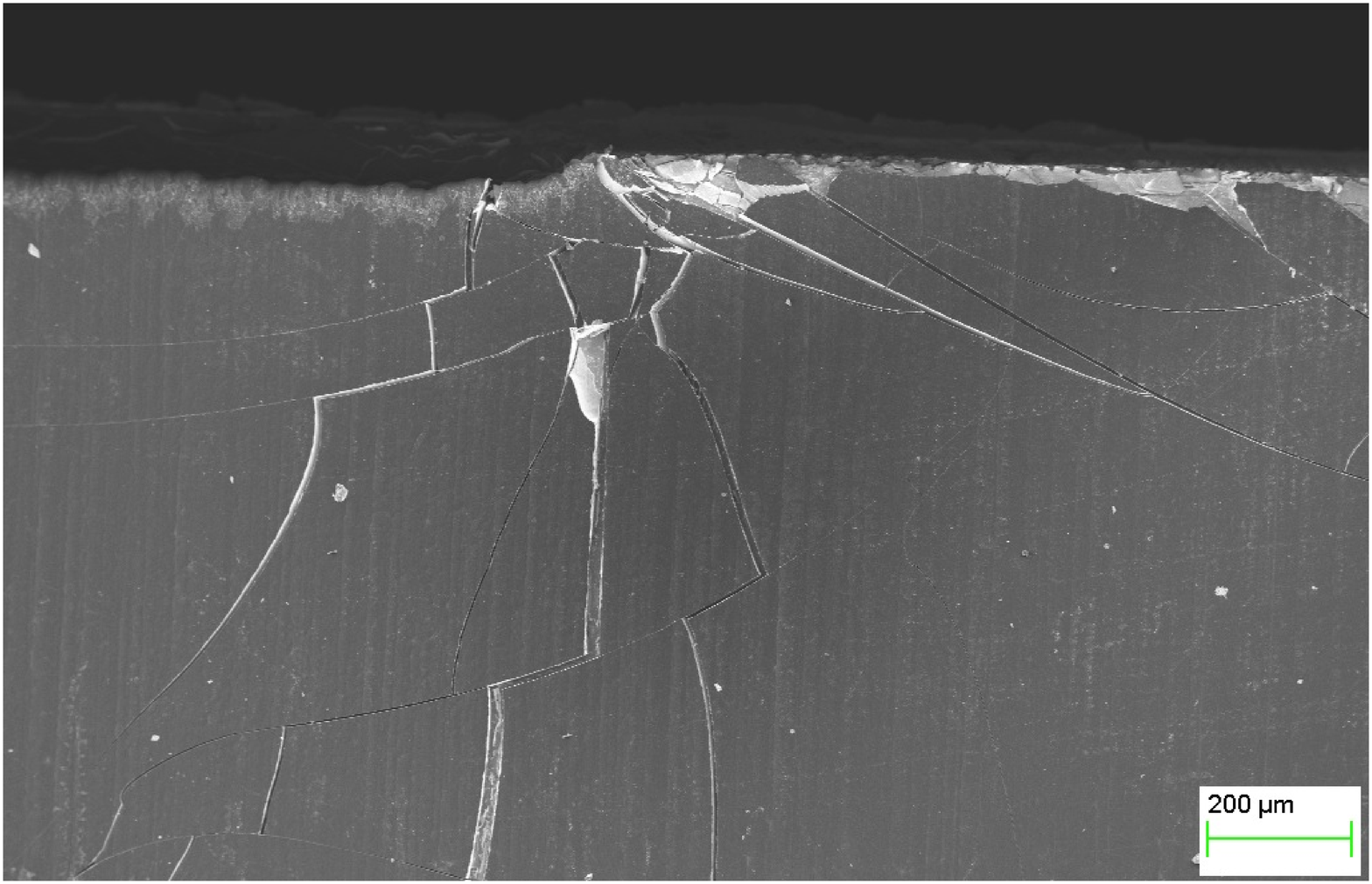}
\includegraphics[width=0.49\textwidth]{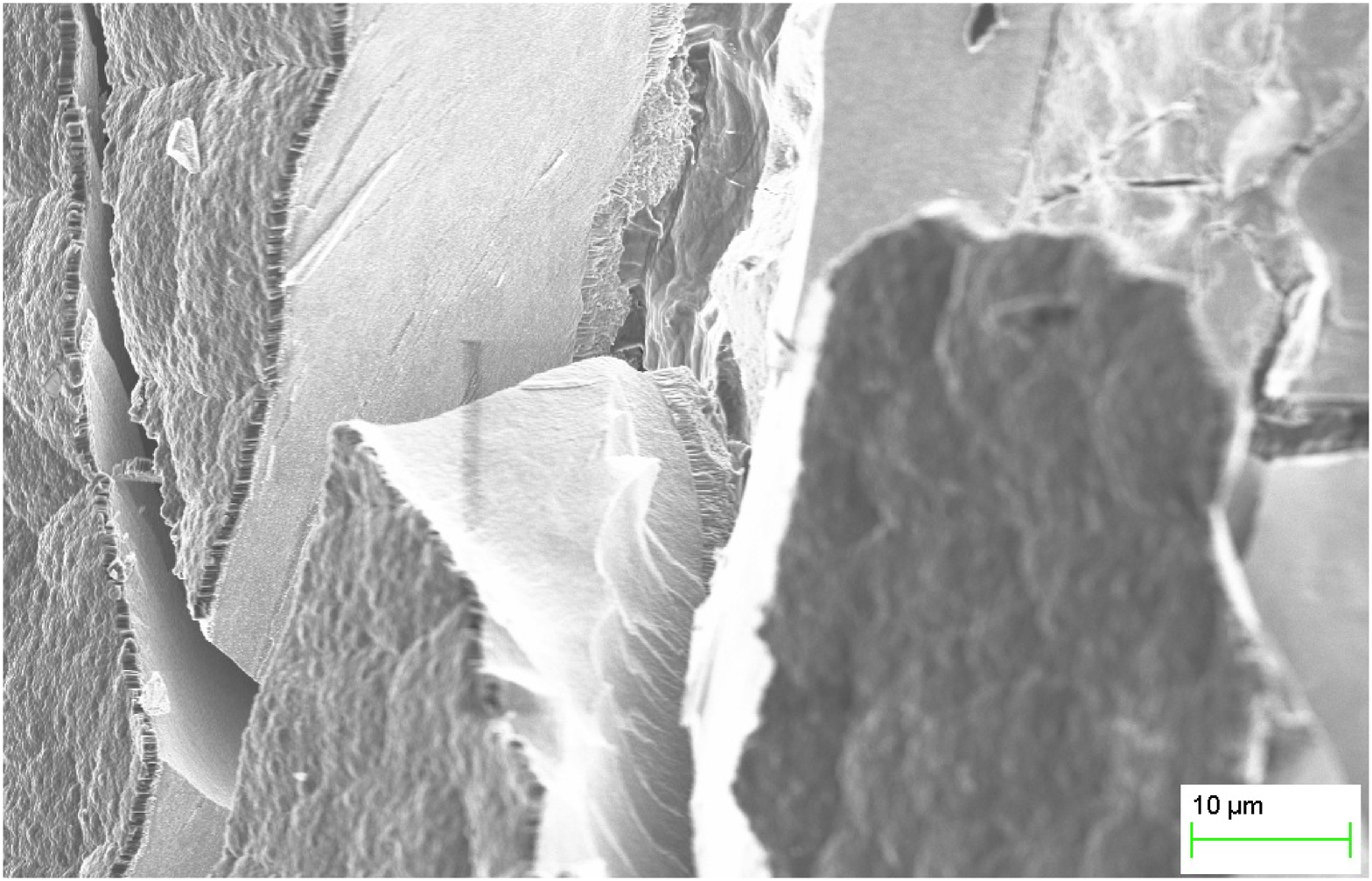} 
\caption{SEM images of cracks in the B$_4$C-coated Ni layer with two magnifications. The location shown is next to an edge of a blade that was cut. The Ni layer (light grey) can be seen to have split and eventually becomes detached from the Al. The B$_4$C layer (dark grey) remains attached to Ni.}
\label{fig:nifail}
\end{figure}

Coating Ni-plated blades with B$_4$C was also attempted. This had initially seemed promising as a good adhesion of B$_4$C to the Ni layer could be achieved. On a long term, however, this solution proved unreliable. Over time (several weeks or more) many blades showed peeling of the Ni layer from the Al. Scanning electron microscopy (SEM) was carried out using a LEO 1550 instrument, equipped with an in-lens detector, operated at 5~kV at a working distance of 2-3~mm, in order to investigate the structure of the surface and coatings and understand the peeling effect. The Ni layer cracks at the edges of blades where they have been cut or stressed. Eventually, cracks in the Ni layer spread over a larger area, see figure~\ref{fig:nifail} (left). It is possible that the process causing this is a gradual access of oxygen to the Al-Ni interface combined with the stress applied during the coating process. This did not occur with any Ni-plated Al parts that were not B$_4$C coated. The image shown in figure~\ref{fig:nifail} (right) is a magnification of the image on the left, close to the edge (although viewed from a different angle). It can be seen here that the adhesion of B$_4$C remains good even when the underlying Ni-layer breaks. 

Interestingly, while, with enough time, blades from all of the three suppliers of chemical plating failed when coated with B$_4$C, the rate was different and possibly correlated to the phosphorous content of the Ni layer. The plating with highest P content, 11\%, failed already during the B$_4$C deposition process. Those with 8\% P, failed days after the deposition. Blades with the lowest P content, 5.5\%, only failed after several weeks or even months, with the majority of the samples still in good condition at the time of the writing.

\section{Conclusion}

Throughout the series of measurements and simulations described here, we have gained a clear understanding of $\alpha$ background in a gaseous neutron detector. While the $\alpha$ activity of typical aluminium, when used in detector parts, is negligible in many measurements, it may be considerable in, for example, Time-of-Flight neutron spectrometers. In our test of the Multi-Grid detector on IN6 at the ILL, this effect elevated the overall background level by at least one order of magnitude, and was the dominant source of background during the experiment. 

We performed measurements of the background in Multi-Grid detectors built using radio-pure aluminium as well as Ni-plated aluminium. With both approaches, we have demonstrated a reduction of the background level by a factor of 50 compared to standard aluminium alloys. With this reduction, the total background in a measurement such as our test on IN6 would be dominated by background due to spurious neutrons, and not $\alpha$ particles. In the case of the Multi-Grid detector, it is practical to Ni-plate the grids in order to maintain a good structural strength, while using radio-pure aluminium for the blades ensuring a good adhesion of the B$_4$C layers. In other types of similar detectors it may be preferable to replace Al entirely by other materials. In case of a detector with a thin coated foil, a nickel foil may be considered. 

We have applied the lessons learned in this study to the construction of our current Multi-Grid demonstrator where 1024 grids were built from Ni-plated and pure aluminium (half-and-half) and all blades, over 18000 in total, were manufactured from pure aluminium and successfully coated with boron carbide.

\acknowledgments

We would like to thank M.~M.~Koza, M.~Zbiri and J.~Halbwachs for the help with measurements at IN6. The work has been supported by the CRISP project (European Commission 7th Framework Programme Grant Agreement 283745).

\end{document}